\documentclass[preprint,showpacs,preprintnumbers,amsmath,amssymb,epsfig]{revtex4}
\usepackage{epsfig,amssymb}
\begin{document}

\title{Relating Neural Dynamics to Neural Coding}
\author{G. Bard Ermentrout${}^{1,3}$, Roberto F. Gal\'an${}^{2,3}$,
  Nathaniel N. Urban${}^{2,3}$}
\affiliation{ ${}^1$University of Pittsburgh, Department of Mathematics, Thackery Hall, Pittsburgh, PA 15260   \\
${}^2$Carnegie Mellon University, Department of Biological Sciences. Mellon Institute, Pittsburgh PA 15213 \\
${}^3$Center for the Neural Basis of Cognition Pittsburgh PA 15213}
\begin{abstract}
 We demonstrate
that two key theoretical objects used widely in Computational
Neuroscience, the phase-resetting curve (PRC) from dynamics and the
spike triggered average (STA) from statistical analysis, are
closely related when neurons fire in a nearly regular manner and the
stimulus is sufficiently small.
We prove  that the STA is proportional to
the derivative of the PRC. We compare these analytic results to numerical
calculations for the Hodgkin-Huxley neuron and we apply the method to
neurons in the olfactory bulb of mice.  This observation
allows us to relate the stimulus-response properties of a neuron to
its dynamics, bridging the gap between dynamical and information
theoretic approaches to understanding brain computations and
facilitating the interpretation of changes in channels and other
cellular properties as influencing the representation of stimuli.
\pacs{  87.18.Sn, 05.45.Xt, 84.35.+i, 89.75.Hc}
\end{abstract}

\maketitle 
Dynamical systems models like the Hodgkin Huxley model of the squid
giant axon are very effective at replicating the firing
properties of individual  neurons.  Such models have been extremely
useful tools for understanding the mechanisms of neural excitability
and for simulating neural circuits.  However, these models
 have been less successful in aiding the
understanding of what neurons compute.  Given
a model that perfectly describes the behavior of a neuron or a
network, we are in most cases still at a loss to say what computation
this neuron or circuit is performing.  Rather, the natural concepts
and objects of the theory of computation (e.g. stimuli, features,
coding) seem only distantly related to those of the theory of
dynamical systems (e.g. differential equations and attractors).  Here
we relate two
mathematical objects central to these two approaches: the phase
resetting curve (PRC) and the spike triggered average (STA).  We start
by introducing these objects and then prove that the STA is
proportional to the derivative of the PRC in the weak stimulus limit.
  We show that this approach allows us to
efficiently and accurately compute the STA from the PRC and vice
versa, in the case of numerical simulations as well as in the case of
real neurons.  The ability to compute these functions from each other
allows us to make some progress in relating dynamics of neurons to
their ability to code features.

The PRC describes how the spiking of a regularly firing neuron is
altered by incoming input, that is, how the time of the next spike is
shifted as a function of the stimulus time relative to the previous
spike: $\Delta(t)\equiv (T-\hat{T}(t))/K$ where $T$ is the natural
period and $\hat{T}(t)$ 
 is the time of the spike given a stimulus at time $t$ after the last
spike. The constant $K$ is proportional to the dimensions of the
measured variable, e.g., in neurons, the voltage. 
\cite{kuramoto,winfree} show that for small stimuli, $x(t)$,
any stable limit cycle oscillator can be reduced to a scalar model for
its phase, $\theta$:
\begin{equation}
\label{eq:1}
\frac{d\theta}{dt} = 1 + \Delta(\theta)x(t)
\end{equation}
where $\theta\in[0,T)$ and  $\Delta(\theta)$ is the PRC. For neurons,
  the stimulus has dimensions of millivolts per millisecond. 
  We take $\theta=0$ to be the time of spiking.  The PRC is valid for
 neurons which are repetively firing (that is, on a stable limit
 cycle), but the concept of the PRC can be applied even when the
 neurons are quite noisy \cite{ger,galan}. $\Delta(\theta)$
 is readily computed for differential equation models, and can also be
 computed for neural models either via direct
  perturbation\cite{rf}-\cite{netoff} or indirectly \cite{ger,galan}.
 Equation (\ref{eq:1}) shows that for a regularly firing neuron, the
 PRC provides, an answer to the question of ``how will a stimulus
 influence when the  next spike will come?''

A contrasting approach to understanding neural computation has focused
on neural coding, by which we mean determining what features of
stimuli are represented by single spikes and spike trains
\cite{spikes}.  
Such
analysis of the stimulus dependence of spike trains often includes the
calculation of the STA, which is related to
the reverse correlation \cite{spikes,DA}.   
The STA  is 
defined as the average stimulus with a given prior statistics 
preceding an action potential in a
neuron.  For our purposes, by “stimulus” we refer to the current
injected into a neuron (divided by the capacitance).  
In other experimental protocols, the stimulus may be a sensory
stimulus presented to an animal while a neuron is recorded. If $x(t)$ is the stimulus:
\begin{equation}
\label{eq:2}
\mbox{STA}(t)= \left<x(\tau_j-t)\right>
\end{equation}
where the average is taken over all spike times, $\tau_j.$  
In the present context, the STA is an answer to the question of ``what temporal features of the
stimulus lead to spiking?'' 

\bigskip
{\it Theory-} 
To derive the main result which applies when the noise is {\em small},
we write $x(t)=\sigma \xi(t)$ where $\sigma$ is the magnitude of the
signal and $\xi(t)$ has, e.g. unit variance. (Note that for white
noise applied to the voltage equation for a neuron model,
 $\sigma^2$ has dimensions of $\mbox{mV}^2/\mbox{ms}.$)
We will use the smallness
of $\sigma$ to estimate the time of a spike in equation (\ref{eq:1})
and from this obtain the STA from (\ref{eq:2}). With $\theta(0)=0$ as
the initial condition (as we assume the neuron has just spiked) we write
\[
\theta(t)=\theta_0(t)+\sigma \theta_1(t) + \ldots 
\]
and substitute into (\ref{eq:1}). Clearly  $\theta_0(t)=t$ and from
this, we find that 
\[
\frac{d\theta_1}{dt} = \Delta(t)\xi(t)
\]
which upon integration yields
\[
\theta_1(t) = \int_0^t \Delta(s)\xi(s)\ ds.
\]
We want to determine the time $\tau$ at which the oscillator spikes
again, that is $\theta(\tau)=T.$ As with $\theta$, $\tau$ depends on
$\sigma$, so we write $\tau=\tau_0+\sigma \tau_1+\ldots$, and find
$\tau_0=T$ and 
\[
\tau_1=-\int_0^T \Delta(s)\xi(s)\ ds.
\] 
Thus, we find that to order $\sigma$,
\begin{equation}
\label{eq:tau}
\tau = T - \sigma \int_0^T \Delta(s)\xi(s)\ ds.
\end{equation}
Substituting  equation (\ref{eq:tau}) into (\ref{eq:2}) and taking
expectations, we get:
\begin{eqnarray}
\mbox{STA}(t) &=& \sigma \left<\xi\left(T-\sigma \int_0^T \Delta(s)\xi(s)\
ds -t\right)\right> \nonumber \\
{} &=& \sigma \left<\xi(T-t)\right> \nonumber \\
{} &{-}&  \sigma^2
\left<\xi'(T-t)\int_0^T\Delta(s)\xi(s)\ ds\right> + \ldots \nonumber \\
{} &=& -\sigma^2 \int_0^T \Delta(s)\left<\xi'(T-t)\xi(s)\right>\ ds +
\ldots \nonumber \\
{} &=& -\sigma^2 \int_0^T \Delta(s) C'(T-t-s)\ ds + \ldots \nonumber
\\
\label{eq:staf} 
{} &=& -\sigma^2 \int_0^T \Delta'(s) C(T-t-s)\ ds + \ldots  
\end{eqnarray}
where $C(t)=\left<\xi(t)\xi(0)\right>$ is the auto correlation of the stimulus. To
get the fourth line, we note that the expected value of $\xi(t)$ is
zero; to get the last line, we have integrated by parts. Dropping the
higher order terms and assuming white noise, we obtain the main
result:
\begin{equation}
\label{eq:staprc}
\mbox{STA}(t) = -\sigma^2 \Delta'(-t)
\end{equation}
where we use the $T-$periodicity of $\Delta(t)$ to drop the $T.$
This result shows that the dynamics of a neuron, as captured by the
PRC, can be used to predict the STA and conversely, given the STA,
we can estimate the PRC of a repetitively firing
neuron. Furthermore, (\ref{eq:staf}) is valid for any zero-mean, stationary, stochastic stimulus.

Two key issues arise in this analysis.  First, we must translate the
periodic PRC to the generally aperiodic STA.
This conversion is most
straightforward when neuron is firing at a nearly
constant rate and the PRC is well-defined. 
  While the STA is in principle aperiodic, in
reality it is only sensible to define the STA over the time interval
prior to a spike in which there are no other spikes.  Thus the time
over which both the STA and the PRC can be clearly defined is the
interval between spikes, i.e. the average period. 
 Second, we must note that because the PRC is the integral of the
STA, it is defined only up to an additive constant term.  However, if we assume
that the PRC vanishes at $t=0,T$ (as is common in neurons), 
then we can determine the integration constant.

\begin{figure}
\includegraphics[width=3.25in]{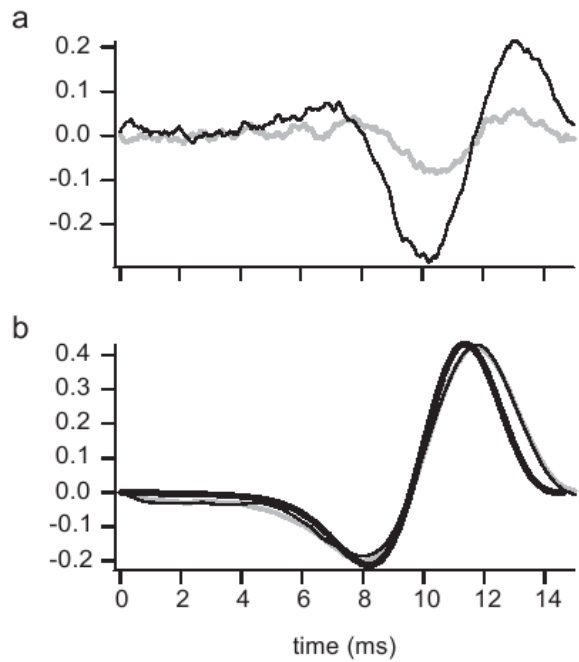}
\caption{Estimation of the PRC from the STA for the Hodgkin-Huxley
  model.(a) STA for two levels of noise (grey: $\sigma=0.25$; black:
  $\sigma=1.0$) (b) 
  PRCs reconstructed from the STA (thick line, true PRC; thin lines,
  reconstructed PRCs)}
\label{fig:f2}
\end{figure}

\bigskip
{\it Examples-}
To test the theory, we compute the STA for the Hodgkin-Huxley (HH)
equations and then use equation (\ref{eq:staprc}) to compute the PRC
subject to the constraints that it vanishes at $t=0$ and at the
average period, $t=T.$ 
 We drive the four variable biophysical
Hodgkin-Huxley model with a constant bias current ($I=10$)to make it fire at 70 Hz, inject noise and
compute the spike triggered average. We then numerically integrate
the STA and time-reverse it to reconstruct an approximate PRC. We compute the
exact PRC using the method in \cite{re}  
and compare the two methods
for two different amplitudes of noise. 
  Figure \ref{fig:f2} shows that in both the low and high noise case, the
 PRCs calculated from the STAs are almost identical to the actual
 PRC. Later (see figure 3), we will systematically quantify the
 dependence of the reconstruction on the statistics of the spike
 trains for both the HH and phase models.
 
   

\begin{figure}
\includegraphics[width=3.25in]{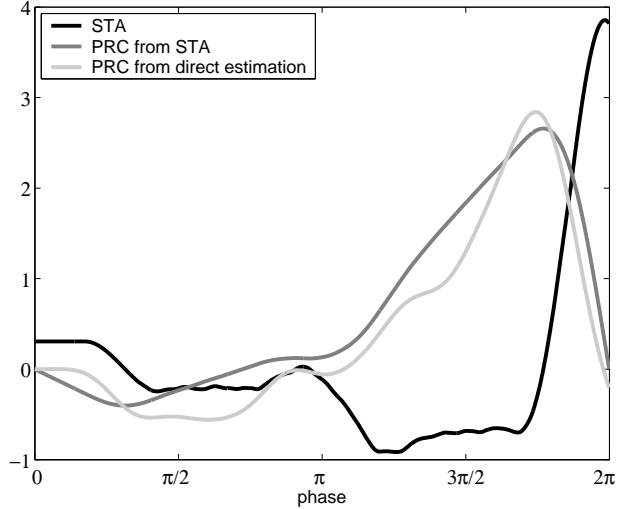}
\caption{Estimation of the PRC of a neuron recorded {\em in vitro}.  Graph
  shows the STA (black) and the PRC estimated using two different methods from
  recordings of an olfactory bulb mitral cell. The light gray
  line shows estimate from method described previously \cite{galan}. The dark gray  line
  shows average PRC for the same cell calculated from the STA, as
  described above}
\label{fig:f3}
\end{figure}

We next tested this transformation on real neurons.  We performed
whole cell recordings from olfactory bulb mitral cells.   We injected these cells
with DC current to cause them to fire repetitively  at $50 \pm6$ Hz,
added noisy current
(with an amplitude 10\% of the DC) and recorded spike times
over intervals of 2-2.5 seconds, repeated 100-120 times.  From the
recorded spike trains we first eliminated the initial period
(250-600 ms) of spike frequency accommodation and then calculated the
STA. 
We calculated the PRC from the STA as per eq. (\ref{eq:staprc}).  For
comparison, we also calculated the PRC using a method based on
injection of aperiodic perturbing pulses \cite{galan}. 
  Figure \ref{fig:f3}  shows the estimated PRC
obtained from these two methods from the an olfactory bulb mitral
cell.  In the PRCs estimated by both methods, there is a substantial
negative region after the spike followed by a larger positive region,
consistent with our earlier estimation of the PRC for olfactory bulb
mitral cells.  We also were interested
in the fact that the STA-based method was possible despite the fact
that cells were not firing in a precisely oscillatory manner.  We measured the standard deviation of the interspike intervals in our
recordings and found it to be approximately 10\% of the firing rate.
Thus our method is robust for at least this level of variation in the
firing rate. We also applied these methods to recordings of neurons 
in the mouse somatosensory cortex (data not shown) with similar results.  

\begin{figure}
\includegraphics[width=3.5in]{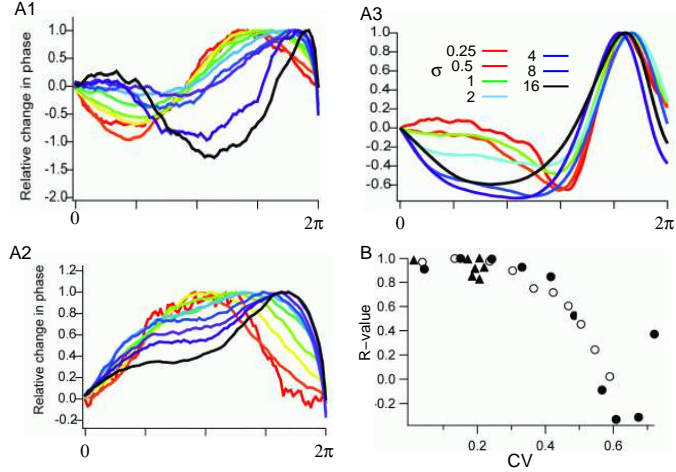}
\caption{Effect of irregular firing on the estimate of the PRC from
  the STA for phase models and the HH model.   Results of
  simulations of phase model with $\Delta(t)=\sin t$ (A1) and
  $\Delta(t)=1-\cos t$ (A2)
  with increasing noise amplitude from 0.2 (red)-2.8 (black)
  in steps of 0.3 (in the order of colors in the rainbow).  For both
  PRCs increasing noise amplitude causes an increase in firing rate.  
 In both cases low noise levels (orange
  traces) produce most precise estimates of the PRCs. (A3)
  Reconstruction of the HH PRC with increasing noise values ($\sigma$
  shown in figure) (B) Plot
  of correlation between actual and estimated PRCs vs CV. (Hollow
  circles, $1-\cos(t)$, solid circles $\sin(t)$ and triangles, HH.) } 
\label{fig:f4}
\end{figure}

To explore further how this relationship between the PRC and the STA
depends on the regularity of the  periodic firing, 
we drive both the HH model and simple phase models of the form of
(\ref{eq:1}) with larger and larger amplitude noise to
make their firing more irregular.  We examined the effects of noise
on phase models with two commonly used PRCs ($\Delta(t)=\sin
t$, Figure \ref{fig:f4}A1; $\Delta(t)=1-\cos t$, Figure \ref{fig:f4}A2).  
  Calculating the STA and the PRC for
these higher noise simulations (after correcting for the change in
average firing rate) resulted in PRCs that less and less closely
resembled the actual PRC for these models although the general shape
of the PRC was somewhat maintained (Figure \ref{fig:f4}A1 and A2). We
injected increasingly larger noise  ($\sigma=0.25-16$) 
into the HH model. While the shape of the PRC
is degraded, the zero crossing is preserved remarkably well.  
We quantified
this degradation in the quality of the estimated PRC by calculating
the correlation coefficient (R) between the actual PRC and the PRC
estimated from the STA and plotted this against the CV of the ISI
distribution.  We observed that for both phase oscillator models the
estimate provided a good approximation (R$>$0.75) of the PRC up to CV =
0.4.  For CVs above 0.4, the correlation between the actual and
estimated PRC declined rapidly for both models. For the HH model, it
was difficult (even with the large noise applied) to increase the CV
beyond about 0.3. Trying larger noise values, led to numerical
difficulties.  We remark that in figure \ref{fig:f4}, we have
normalized the magnitude of the PRC to a maximum of 1 and that the
un-normalized reconstructed PRCs for the HH model had amplitudes that
were smaller at large noise values.   

\bigskip
{\it Discussion-} Relating neural dynamics to neural coding has been termed a grand challenge
for neuroscience and we believe that our work describes an important step
towards relating these two subjects, albeit for a restricted set of stimuli.
Strengthening this
connection will provide valuable means of relating the wealth of data on
biophysical properties of neurons (as captured in dynamical systems models
of neuronal properties) to questions about the properties of stimuli that
are being computed by neurons. The main limitation to our approach, which
still allows it to be applied to many situations, arises from 
 the fact that the PRC, as
useful as it is,
has limits to its applicability; specifically, in this context, it is
defined only for nearly periodically firing neurons where only the
timing of spikes  (and not the rate) are altered.  Nonetheless, neurons in
many brain networks are active spontaneously and the strength of any one
input is weak.  Thus, the assumption that inputs modulate the timing of
spikes rather than adding more spikes may hold.  For example, 
\cite{preyer} demonstrated that realistic synaptic conductances in the aplysia satisfy
the mathematical criteria of ``weak coupling'' in the sense that the notion of phase still makes sense.

\cite{arcas}  were among the first to try to extract
statistical information such as the STA and the spike triggered
covariance (STC)  from a biophysical model
for a neuron. Using simulations of the Hodgkin-Huxley equations, they
compute both the STA and the STC in a situation
where the stimulus is sufficiently strong to elicit spikes. They show that
there are important features of the stimulus encoded in the STC which
are not evident in the STA. The calculations in section II could be
extended to incorporate higher-order effects and thus relate the STC
to other aspects of the PRC \cite{ed}.  \cite{arcas} also consider the effects
of the amplitude of the stimulus on the STA; such effects would be
reflected in changing the shape of the reconstructed PRC (as in figure
\ref{fig:f4}A3). While our theory is linear (with respect to the
effects of the stimulus), it is known that the shape of the PRC is
also affected by stimuli which are sufficiently large \cite{winfree},
so that some of the shape effects in figure \ref{fig:f4} may be due to
pushing the simulations beyond the linear range.   
More recently, \cite{paninski} derived
the STA for the integrate-and-fire model and produced formulae in the limit
of small noise.   \cite{badel} characterized the spike triggered average
voltage (a related quantity which requires intracellular recording of the
membrane voltage fluctuations) for several different models and related it
to the underlying dynamics. 
\cite{hg} compute the PSTH and STA of a general class
of spike-response models, thus providing some insight into the
relationships between dynamics of spiking and the neural code. This
work is the closest to ours in its generality. They apply their
results to the data in \cite{poliakov}.
  Our methods, while in a more restricted
situation (regularly firing) are very general in that they apply to any
model of an experimental system for which a PRC is defined. In
particular, PRCs encode dynamic information about subthreshold
behavior  \cite{ger, prc}, which is not possible in spike-response
models. The PRC can be directly computed from any biophysical model
for periodic neuronal firing.

In this paper, we have described a
relationship between the first order statistics (spike triggered average)
and the PRC.  Because dynamical systems and coding-based approaches are both
concerned with the fundamental question of what makes a neuron spike we
anticipate that further analysis will allow this approach to be extended to
cover situations in which the average firing rate of a neuron is rapidly
varying.

The authors were supported by NSF grants, DC005798, MH079504 and NSF
DMS0513500 and they would like to thank Brent Doiron and Nicolas
Fourcaud-Trocme for their comments on previous versions of this
manuscript.   

\end{document}